\documentclass[aps,prb,twocolumn,showpacs,amsmath,amssymb,superscriptaddress]{revtex4}
\usepackage{dcolumn}
\usepackage{bm}
\usepackage{graphicx}
\usepackage{epstopdf}
\begin{document}

\renewcommand{\section}[1]{{\par\it #1.---}\ignorespaces}

\title{Proximity-induced unconventional superconductivity in topological insulators}
\author{Annica M. Black-Schaffer}
 \affiliation{Department of Physics and Astronomy, Uppsala University, Box 516, S-751 20 Uppsala, Sweden}
 \author{Alexander V. Balatsky}
 \affiliation{Nordic Institute for Theoretical Physics (NORDITA), Roslagstullsbacken 23, S-106 91 Stockholm, Sweden}
 \affiliation{Theoretical Division and Center for Integrated Nanotechnologies, Los Alamos National Laboratory, Los Alamos, New Mexico 87545, USA}
 \date{\today}

\begin{abstract}
We study and classify the proximity-induced superconducting pairing in a topological insulator (TI)-superconductor (SC) hybrid structure for SCs with different symmetries. The Dirac surface state gives a coupling between spin-singlet and spin-triplet pairing amplitudes as well as pairing that is odd in frequency for $p$-wave SCs.
We also find that all SCs induce pairing that is odd in both frequency and orbital (band) index, with  oddness in frequency and orbital index being completely interchangeable. The different induced pairing amplitudes significantly modify the density of states in the TI surface layer.
\end{abstract}
\maketitle

%
Topological insulators (TIs) represent a new class of quantum matter, with a gapless surface state inside the bulk energy gap.\cite{Hasan10, Qi11} The surface state has a Dirac-like energy spectrum with spin and momentum locked into a spin-helical structure. TIs are thus ideally suited to study the combination of Dirac physics with different broken symmetry order parameters, such as magnetism or superconductivity.
Most notably, superconductivity has already been demonstrated in TIs through the proximity effect for both conventional $s$-wave\cite{Sacepe11, Veldhorst12} and high-temperature $d$-wave cuprate superconductors (SCs).\cite{Zareapour12}

The low-energy spectrum of a TI proximity-coupled to a spin-singlet $s$-wave SC resembles that of a spinless $p+ip'$-wave SC.\cite{Fu08} The allure of the spinless $p+ip'$-wave SC is that it hosts Majorana fermions in, e.g., vortex cores and Josephson junctions \cite{Read00, Fu08}, and TI-SC hybrid structures have therefore lately received a lot of attention.\cite{Alicea12, Beenakker11} A Majorana fermion is its own anti-particle and obeys non-Abelian statistics, which supports fault-tolerant quantum computation.\cite{Nayak08}
Aside from the simple induction of conventional order, in a spatially varying superconducting state, the TI surface state also contains odd in time, or odd-frequency, superconducting pairing.\cite{Black-Schaffer12oddw}

These results indicate a possibly complex set of induced pairing amplitudes in a TI-SC hybrid structure.
In this Rapid Communication, we provide a full symmetry classification of all induced pairing amplitudes for a TI proximity-coupled to SCs with spin-singlet $s$- and $d$-wave, as well as spin-triplet $p$-wave symmetries.
We not only classify the coupling between spin-singlet and spin-triplet states for all different SC symmetries, but also show that odd-frequency (intraorbital) pairing appears for spin-triplet SCs. Moreover, for all SCs, we find pairing that is both odd in frequency and odd in orbital (or band) index. In fact, we find a complete reciprocity between pairing that is odd in orbital index and odd in frequency. The induced pairing amplitudes are reflected in the local density of states (LDOS) of the TI surface state.
The couplings between different pairing symmetries are not restricted to proximity superconductivity into the surface state but extend to the whole TI, independent on doping level, and this shows on an intricate mixture of different pairing symmetries in TIs.

%
 As a prototype TI we use Bi$_2$Se$_3$, modeled by its two Bi orbitals on a cubic lattice with side $a$:\cite{Rosenberg12}
%
\begin{align}
\label{eq:HTI}
H_{\rm TI}  =   \gamma_0 - 2\sum_{{\bf k},i} \gamma_i \cos(k_i a) + \sum_{{\bf k},\mu} d_\mu \Gamma_\mu.
\end{align}
Here, $d_0 = \epsilon - 2\sum_i t_i \cos(k_i a)$, $d_i = -2\lambda_i \sin(k_i a)$, $\Gamma_0 = \tau_x \otimes \sigma_0$, $\Gamma_x = -\tau_z \otimes \sigma_y$, $\Gamma_y = \tau_z \otimes \sigma_x$, and $\Gamma_z = \tau_y \otimes \sigma_0$, with $\tau_i$ and $\sigma_i$ being the Pauli matrices in orbital and spin space, respectively.
The parameters, fitted to the Bi$_2$Se$_3$ dispersion at the $\Gamma$-point,\cite{Zhang09, Li10, Rosenberg12} are $\gamma_0 = 0.3391$, $\gamma_{x,y} = 0.0506$, $\gamma_z = 0.0717$, $\epsilon = 1.6912$, $t_{x,y} = 0.3892$, $t_z = 0.2072$, $\lambda_{x,y} = 0.2170$, and $\lambda_z = 0.1240$~eV.
The Bi orbitals are shifted away from the inversion center in the $z$-direction and thus do not have definite parity, which is necessary when using the parity of interorbital pairing to classify the induced superconducting amplitudes.
%
\begin{table*}[t]
    \footnotesize
  \begin{tabular}{| c  | c |  c || c | c | c | c |}
 \hline
 \multicolumn{3}{| c ||}{Superconductor} & \multicolumn{2}{ c |}{Even-frequency} & \multicolumn{2}{ c |}{Odd-frequency}\\
 \hline
 \multicolumn{1}{|c|}{$\Gamma$} & \multicolumn{1}{c|}{Basis function} & \multicolumn{1}{c|| }{$J_z$} & \multicolumn{1}{c| }{Even-orbital} & \multicolumn{1}{c |}{Odd-orbital} & \multicolumn{1}{ c |}{Even-orbital} & \multicolumn{1}{ c |}{Odd-orbital} \\
  \hline \hline
 A$_{\rm 1g}$ & $\psi = 1$ & 0 & A$_{\rm 1g}$ singlet, & - & - & A$_{\rm 1g}$ singlet, \\
  & & &  A$_{\rm 2u}$ triplet (m$_s = \pm 1$) &  &  & A$_{\rm 2u}$ triplet (m$_s = \pm 1$) \\
  \hline
 B$_{\rm 1g}$ & $\psi = k_x^2-k_y^2$ & $\pm 2$ & B$_{\rm 1g}$ singlet, & - & - & B$_{\rm 1g}$ singlet,\\
 & & &  B$_{\rm 2u}$ triplet (m$_s = \pm 1$) & &  & B$_{\rm 2u}$ triplet (m$_s = \pm 1$) \\
 \hline
 B$_{\rm 2g}$ & $\psi = 2k_x k_y$ &  $\pm 2$ & B$_{\rm 2g}$ singlet, & - & - & B$_{\rm 2g}$ singlet,\\
 & & &  B$_{\rm 1u}$ triplet (m$_s = \pm 1$) &  &  & B$_{\rm 1u}$ triplet (m$_s = \pm 1$) \\
 \hline
 \hline
 A$_{\rm 1u}$ & ${\bf d} = (k_x,k_y,0)$ & 0 & A$_{\rm 1u}$ triplet (m$_s = \pm 1$) & A$_{\rm 1g}$ triplet  (m$_s = 0$) & A$_{ \rm1g}$ triplet  (m$_s =  0$) & A$_{\rm 1u}$ triplet (m$_s = \pm 1$)  \\
  \hline
A$_{\rm 2u}$ & ${\bf d} = (k_y,-k_x,0)$ & 0 & A$_{\rm 2u}$ triplet (m$_s = \pm 1$), & - & - & A$_{\rm 2u}$ triplet (m$_s = \pm 1$),  \\
 & & &  A$_{\rm 1g}$ singlet &  &  & A$_{\rm 1g}$ singlet \\
  \hline
 B$_{\rm 1u}$ & ${\bf d} = (k_x,-k_y,0)$ & $\pm 2$ & B$_{\rm 1u}$ triplet (m$_s = \pm 1$), & B$_{\rm 1g}$ triplet  (m$_s = 0$) & B$_{\rm 1g}$ triplet  (m$_s =  0$) & B$_{\rm 1u}$ triplet (m$_s = \pm 1$),  \\
 & & &  B$_{\rm 2g}$ singlet &  &  & B$_{\rm 2g}$ singlet \\
  \hline
 B$_{\rm 2u}$ & ${\bf d} = (k_y,k_x,0)$ & $\pm 2$ & B$_{\rm 2u}$ triplet (m$_s = \pm 1$), & B$_{\rm 2g}$ triplet  (m$_s = 0$) & B$_{\rm 2g}$ triplet  (m$_s =  0$) & B$_{\rm 2u}$ triplet (m$_s = \pm 1$),  \\
 & & &  B$_{\rm 1g}$ singlet &  &  & B$_{\rm 1g}$ singlet \\
 \hline
 E$_{\rm 2u}^+$ & ${\bf d} = (0,0,k_x+ik_y)$ & $1$ & E$_{\rm 2u}^+$ triplet (m$_s = 0$) & A$_{\rm 1g}$ triplet  (m$_s = 1$), & A$_{\rm 1g}$ triplet  (m$_s =  1$), & E$_{\rm 2u}^+$ triplet (m$_s = 0$)  \\
 & & &  & B$_{\rm 1g} \!+\! i$B$_{\rm 2g}$ triplet (m$_s = -1$) & B$_{\rm 1g} \!+\! i$B$_{\rm 2g}$ triplet (m$_s = -1$) & \\
  \hline
E$_{\rm 2u}^-$ & ${\bf d} = (0,0,k_x-ik_y)$ & $-1$ & E$_{\rm 2u}^-$ triplet (m$_s = 0$) & A$_{\rm 1g}$ triplet  (m$_s = -1$), & A$_{\rm 1g}$ triplet  (m$_s =  -1$), & E$_{\rm 2u}^-$ triplet (m$_s = 0$)  \\
 & & &  & B$_{\rm 1g} \!-\! i$B$_{\rm 2g}$ triplet (m$_s = 1$) & B$_{\rm 1g} \!-\! i$B$_{\rm 2g}$ triplet (m$_s = 1$) & \\
  \hline
 \end{tabular}
 \caption{Proximity-induced superconductivity in a Bi$_2$Se$_3$-type TI with tetragonal symmetry. The 2D SCs are classified into the irreducible representations $\Gamma$ of the D$_{4h}$ group ($k_z = 0$ and ignoring the $g$-wave A$_{\rm 2g}$ representation), with basis functions $\psi$ and ${\bf d}$ [on the square lattice $k_i \rightarrow \sin(k_ia), k^2_i \rightarrow 2(1-\cos(k_ia))$] and total angular momentum $J_z$.
The proximity-induced pairing amplitudes are classified into even- and odd-frequency, even-orbital (intraorbital and even-interorbital) and odd-orbital (odd-interorbital) components. The magnetic quantum number m$_s$ is given for all spin-triplet amplitudes.}
 \label{tab}
\end{table*}
%
We create a 20 layer thick slab of $H_{\rm TI}$ in the $z$-direction and add a SC to the top surface. We define the SC on a square lattice with side $a$:
%
\begin{align}
\label{eq:HSC}
H_{\rm SC}  & =  \sum_{{\bf k},\sigma} \varepsilon({\bf k}) c_{{\bf k}\sigma}^\dagger c_{{\bf k}\sigma} \ \ + \\ \nonumber
& \frac{1}{2}\sum_{{\bf k},\sigma,\sigma'} \left[ \Delta_{\sigma \sigma'}({\bf k}) c^\dagger_{{\bf k}\sigma} c^\dagger_{-{\bf k}\sigma'} - \Delta_{\sigma \sigma'}^\ast (-{\bf k}) c_{-{\bf k}\sigma} c_{{\bf k}\sigma'} \right],
\end{align}
where $c_{{\bf k}\sigma}^\dagger$ creates an electron with momentum ${\bf k} = (k_x,k_y)$, spin $\sigma$, and $\varepsilon({\bf k}) = -2[\cos(k_x a) + \cos(k_y a)] + \mu_{\rm SC}$. The superconducting order parameter can be written as $\hat{\Delta}({\bf k}) = i \Delta_0 \sigma_y \psi({\bf k})$ for spin-singlet pairing and $\hat{\Delta}({\bf k}) = 2i\Delta_0({\bf d}({\bf k})\cdot {\bm \sigma}) \sigma_y$ for spin-triplet pairing,\cite{Sigrist91}$^,$\footnote{This definition make spin-singlet and spin-triplet pairing on nearest neighbor bonds equivalent.} with $\Delta_0$ the pairing gap and basis functions $\psi$ and ${\bf d}$ given in Table \ref{tab}.
Finally, we couple the SC and the TI with a local tunneling Hamiltonian:
%
\begin{align}
\label{eq:Ht}
H_{\tilde{t}}  & =  -\sum_{{\bf k},\sigma} \tilde{t}_1 c^\dagger_{{\bf k}\sigma}b_{1{\bf k}\sigma} + \tilde{t}_2 c^\dagger_{{\bf k}\sigma}b_{2{\bf k}\sigma} + {\rm H.c.},
\end{align}
where $b^\dagger_{a{\bf k}\sigma}$ creates an electron in orbital $a =1, 2$ in the TI surface layer.
We solve $H = H_{\rm TI} + H_{\rm SC}+ H_{\tilde{t}}$ using exact diagonalization and are here primarily interested in the different time-ordered pairing amplitudes in the TI surface layer:\footnote{The induced superconducting amplitudes are decreasing exponentially with distance from the surface.}
%
\begin{align}
\label{eq:F}
F_{\sigma\sigma'}^{a b}(\tau) = \frac{1}{2N_{\bf k}} \sum_{\bf k} S_{\sigma \sigma'}({\bf k}) \cal{T}_\tau \langle & b_{a -{\bf k}\sigma'}(\tau)b_{b {\bf k}\sigma}(0) \pm  \\ \nonumber
& b_{b -{\bf k}\sigma'}(\tau)b_{a {\bf k}\sigma}(0)\rangle,
\end{align}
with even ($+$) and odd ($-$) pairing in orbital index and $N_{\bf k}$ being the number of ${\bf k}$-points in the Brillouin zone.
$F_{\sigma\sigma'}^{a b}$ can also either be even or odd in time ($\tau$), or equivalently frequency ($\omega$). The even-frequency pairing amplitude is the equal-time amplitude $F_{\sigma\sigma'}^{a b}(\tau = 0)$. For the odd-frequency pairing, amplitude we use the time derivative at equal times $\partial F_{\sigma\sigma'}^{a b}(\tau)/\partial \tau | _{\tau = 0}$, which is only non-zero for odd-time dependence.\cite{Balatsky93, Abrahams95, Dahal09, Black-Schaffer12oddw}
The symmetry factor $S_{\sigma \sigma'} = \Delta_{\sigma \sigma'}^\ast/\Delta_0$ for even-frequency even-orbital (intraorbital and even-interorbital)  or odd-frequency odd-orbital (odd-interorbital) pairing. For pairing odd in the orbital index or frequency, Fermi statistics requires spin-singlet amplitudes to have an odd-${\bf k}$ $S$ factor ($p$-wave) and spin-triplet states to have an even-${\bf k}$ $S$ factor ($s$- or $d$-wave).

In Table \ref{tab} we list all the proximity-induced pairing amplitudes for the physically relevant 2D SC symmetries in the D$_{4h}$ group.\cite{Sigrist91}
First we focus on the {\it even-frequency even-orbital} amplitudes, where Fermi statistics gives the usual spin-singlet even-${\bf k}$ or spin-triplet odd-${\bf k}$ combinations. Naturally, the primary amplitude, i.e.~of the symmetry of the SC, is always found among the TI pairing amplitudes. In addition, the spin-momentum locking in the Dirac surface state has been shown to induce a $p$-wave state for a spin-singlet $s$-wave SC.\cite{Fu08, Stanescu10, Black-Schaffer11QSHI,Lababidi11, Nandkishore13, Tkachov13} Here we are able to further classify this amplitude as a spin-triplet A$_{\rm 2u}$ state.
The appearance of A$_{\rm 2u}$ pairing instead of A$_{\rm 1u}$ is due to the effective low-energy Dirac surface state Hamiltonian $H_{\rm TIsurf} = \sum_{\bf k} v(k_x\sigma_y - k_y \sigma_x)$, with $v$ being the Fermi velocity.\cite{Rosenberg12} We also find that a spin-singlet $d$-wave SC similarly induces a spin-triplet B$_{\rm 1u/2u}$ state, due to conservation of total angular momentum $J_z$ (the rotational symmetry around the $z$-direction is assumed to be an intact symmetry).
By reciprocity, spin-triplet $p$-wave SCs necessarily also induce the corresponding spin-singlet even-${\bf k}$ amplitudes.
Using a spinless linear combination of the spin-full surface state operators, $H_{\rm TIsurf}$ with spin-singlet $s$-wave and spin-triplet A$_{\rm 2u}$ superconducting pairing can be written as an effective spinless $p_y+ip_x$-wave state, which supports Majorana fermions.\cite{Read00, Fu08}
Although a $d$-wave SC also induces an equal-spin triplet state, a similar procedure does not yield a simple spinless $p+ip'$ superconducting state.

Next we discuss the {\it odd-frequency even-orbital} amplitudes in Table \ref{tab}. Numerically, we find that no spin-singlet SC induces such amplitudes, but they are in general present for spin-triplet SCs. Oddness in frequency invokes a change from odd to even momentum parity, keeps the spin-triplet nature, but modifies the magnetic quantum number m$_s$ in order to preserve $J_z$.
It is possible to show also analytically that spin-triplet $p$-wave SCs induce odd-frequency intraorbital pairing in a Dirac system, such as the TI surface state.
For this purpose, we start with the effective low-energy (single orbital) Hamiltonian $H_{\rm TI surf} = \sum_{\bf k} v (\tilde{{\bf k}}\cdot {\bm \sigma})$, with $\tilde{\bf k} = (-k_y,k_x,0)$,\cite{Rosenberg12} coupled to a spin-triplet $p$-wave SC through a local tunneling element $\tilde{t}$.
The anomalous pairing propagator induced in the TI is $\hat{F}_{\rm TI}({\bf k}, \omega_n)  = |\tilde{t}|^2 \hat{G}({\bf k}, \omega_n) \hat{F}_s({\bf k}, \omega_n)\hat{G}({-\bf k}, -\omega_n)$, where $\hat{G}({\bf k}, \omega_n) = [i\omega_n - v\tilde{\bf k} \cdot {\bm \sigma}]/[\omega_n^2 + k^2]$ is the normal Green's function in the TI and $\hat{F}_s({\bf k}, \omega_n) = \hat{\Delta}({\bf k})/[\omega_n^2 + E_k^2]$ is the superconducting Green's function with $E_k$ the Bogoliubov quasiparticle energies. Hat symbols represent the spin-matrix structure of the Green's functions.
We will here use a standard perturbation approach\cite{Black-Schaffer12oddw} and only focus on the linear in $\omega_n$ odd-frequency component. For a spin-singlet SC, an odd-frequency component is only present in the TI for a spatially inhomogenous order parameter\cite{Black-Schaffer12oddw, Tanaka07PRL, Tanaka07} or in a magnetic field \cite{Yokoyama12}, but for a homogenous spin-triplet SC, we find an induced odd-frequency component:
%
\begin{align}
\label{eq:ind1}
\hat{F}_{\rm TI}({\bf k}, \omega_n) = \frac{-4iv|\tilde{t}|^2 \Delta_0 [\tilde{\bf k} \times {\bf d}({\bf k})]\cdot{\bm \sigma} \sigma_y}{(\omega_n^2 +k^2)^2(\omega_n^2 + E_k^2)} \omega_n.
\end{align}
The odd-frequency component thus has a spin-triplet structure with an effective
%
\begin{equation}\label{EQ:induced}
{\bf d}_{\rm eff}({\bf k})  = \tilde{\bf k} \times {\bf d}({\bf k}),
\end{equation}
which is an {\it even} function of momentum since ${\bf d}$ is linear in ${\bf k}$. Explicitly working out ${\bf d}_{\rm eff}$, we again arrive at the results in Table I.   
The on-site amplitude of the odd-frequency intraorbital spin-triplet pairing can be calculated as a sum over all momenta of the amplitude in Eq.~(\ref{eq:ind1}). The result depends on the relative strength of the DOS of TI versus SC: (i) for DOS of TI $>$ DOS of the metal $N_0$, we find $\sum_{\bf k} \hat{F}_{TI}({\bf k}, \omega_n) \sim i\Delta_0/(\omega_n E^2_F)$, whereas for (ii) DOS of TI $<$ $ N_0$, $\sum_{\bf k} \hat{F}_{TI}({\bf k}, \omega_n) \sim i N_0 \omega_n$.
The above result also implies that any order parameter with a linear-${\bf k}$ dependence along the TI interface induces (intraorbital) odd-frequency pairing. We have confirmed this numerically for a 2D TI proximity coupled to a $d$-wave SC with a node along the interface.

Finally, we discuss the presence of {\it odd-orbital} amplitudes in Table \ref{tab}. The two Bi orbitals have different distances to the SC, so the occurrence of odd-orbital pairing might not be fully unexpected.
 We find a complete reciprocity between parity in the orbital and frequency domains. Any amplitudes with even-orbital, even-frequency symmetry (column 4 in Table \ref{tab}) are always accompanied by an odd-orbital, odd-frequency amplitude with the same momentum and spin symmetry (i.e., column 5 = column 4). This odd-orbital, odd-frequency pairing does not break time-reversal symmetry unless the even-orbital, even-frequency pairing does so.
 Likewise, any even-orbital, odd-frequency amplitudes (column 6) also come with odd-orbital, even-frequency pairing (column 7). Thus, knowing the content in column 4 and 6, as discussed above, we can completely determine all induced pairing amplitudes.
The complete interchangeability of orbital and frequency symmetries is not only restricted to TIs, but is found generally in two-orbital systems with a finite interorbital hybridization.\cite{Black-Schaffer13new}

The results in Table~\ref{tab} are derived using the tetragonal point group for the TI, but the surface of Bi$_2$Se$_3$ has a hexagonal symmetry, described by the D$_{6h}$ point group. Fortunately, only considering $k_z = 0$ and orbital angular momentum 2 or lower, there is only minor differences between these two groups. The $d$-wave representations B$_{\rm 1g/2g}$ are transformed into the two-dimensional E$_{\rm g}$ representation of D$_{6h}$, and likewise for the B$_{\rm 1u/2u}$ representations. Since these $J_z =\pm 2$ representations only generate pairing amplitudes within themselves, Table I does not change.

\begin{figure}[t!]
\includegraphics[scale = 0.95]{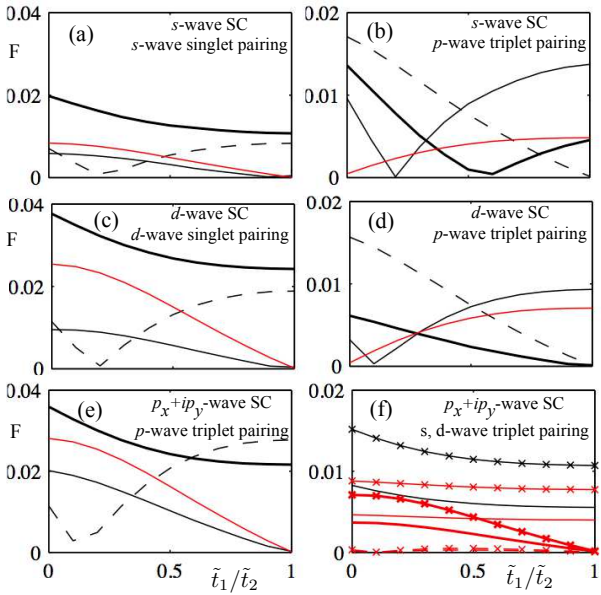}
\caption{\label{fig:OP} (Color online) Proximity-induced pairing amplitudes in the TI surface state from an $s$-wave (a, b), $d$-wave (c,d), and $p_x+ip_y$-wave (e,f) SC as a function of orbital tunneling ratio $\tilde{t}_1/\tilde{t}_2$ for $\tilde{t} = 0.9$, $\Delta_0 = 0.5$, and $\mu_{\rm SC} = -0.5$~eV. Symmetry of the SC amplitudes (a,c,e) and triplet $p$-wave amplitudes (b,d) are divided into even-intraorbital (thick black), odd-intraorbital (thin black), even-interorbital (dashed black), and odd-frequency odd-orbital (thin red) pairing amplitudes. For the $p_x+ip_y$-wave SC the induced spin-triplet $s$-wave (line) and $d$-wave (crosses) amplitudes (f) are divided into odd-orbital (black), odd-frequency even-intraorbital (thick red), odd-intraorbital (thin red), and even-interorbital (dashed red) amplitudes.
}
\end{figure}

In terms of pairing strengths, we find that all amplitudes in Table I increase linearly with the superconducting order parameter $\Delta_0$ and quadratically with the norm of the tunneling amplitude $\tilde{t} = (\tilde{t}_1^2 + \tilde{t}_2^2)^{-1/2}$, for fixed ratio $\tilde{t}_1/\tilde{t}_2$. Moreover, we find only a very weak dependence on the chemical potential in the TI, notably, the amplitudes do not go to zero for $\mu_{\rm TI} = 0$.\footnote{When including the ${\bf k}$-dependent $S_{\sigma \sigma'}(\bf k)$, many amplitudes are still of course zero at the $\Gamma$-point.}
In Fig.~\ref{fig:OP}, we plot the dependence on the orbital tunneling ratio $\tilde{t}_1/\tilde{t}_2$ for $s$-wave (a,b), $d$-wave (c,d), and $p_x+ip_y$-wave (e,f) SCs. For clarity, we have divided the intraorbital pairing into even- and odd-intraorbital pairing, although they are both even functions in orbital-space.
%
For the SC symmetry (left column), we see that both the even- and odd-intraorbital pairing (thick and thin black lines) decreases with increasing $\tilde{t}_1/\tilde{t}_2$ ratio, whereas the even-interorbital pairing (dashed) goes through zero around 0.1. There is thus no significant change in the overall even-orbital amplitude.
The odd-frequency odd-orbital amplitude (red) tracks the odd-intraorbital pairing and is a sizable fraction of the even-frequency pairing.
The spin-triplet amplitudes for spin-singlet SCs, see Figs.\ref{fig:OP}(b,d), also show no significant change in overall even-orbital amplitudes, whereas the odd-frequency component slowly increases with $\tilde{t}_1/\tilde{t}_2$ ratio. The spin-triplet amplitudes can reach up to $80$\% of the spin-singlet amplitudes for an $s$-wave SC, but are somewhat smaller for a $d$-wave SC. For the $k_x + ik_y$-wave SC the induced $s$- and $d$-wave spin-triplet amplitudes, see Fig.~\ref{fig:OP}(f), are only moderately weakly dependent on $\tilde{t}_1/\tilde{t}_2$.
To summarize, all amplitudes in Table I can be significant in size, at least for single orbital dominated tunneling.

\begin{figure}[t!]
\includegraphics[scale = 1.1]{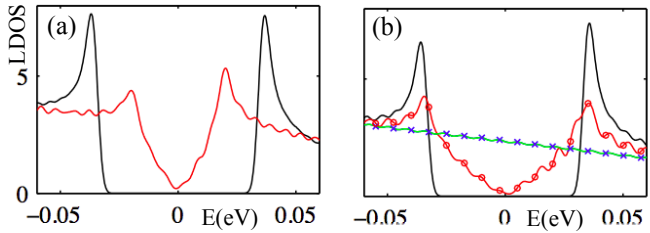}
\caption{\label{fig:LDOS} (Color online) LDOS (states/eV/unit cell) in the TI surface state as a function of energy for a spin-singlet SC (a) with $s$-wave (black) and $d$-wave (red) symmetries and for a spin-triplet SC (b) with $p$-wave A$_{\rm 1u}$ (green line), A$_{\rm 2u}$ (black line), B$_{\rm 1u/2u}$ (red line with circles), and E$_{\rm 2u,+/-}$ (blue crosses) symmetries. Here, $\Delta_0 = 5$, $\mu_{\rm SC} = -0.3$, and $\tilde{t} = (0,0.6)$. The Dirac point is found at higher energies due to doping from the SC. Ripples are due to finite ${\bf k}$-point sampling.}
\end{figure}

The different induced pairing amplitudes have important consequences for the local density of states (LDOS) in the TI surface layer, as displayed in Fig.~\ref{fig:LDOS}.
%
As expected, a spin-singlet $s$-wave SC induces a finite gap in the energy spectrum, while a spin-singlet $d$-wave SC produces a narrow nodal LDOS. The superconducting gap appears at zero energy, whereas the surface state Dirac point in general moves to finite energies due to doping from the SC.
The spin-triplet $p$-wave SCs, on the other hand, cause a wide variety of low-energy LDOS spectra, from fully gapped, to nodal, to no observable consequences of superconductivity. Linder {\it et al.}\cite{Linder10PRL, Linder10PRB} established that any spin-triplet $p$-wave pairing amplitude in a TI only renormalizes the chemical potential and thus never gaps the surface energy spectrum.
However, we find that only the A$_{\rm 1u}$ and E$_{\rm 2u}$ energy spectra remain gapless.
The discrepancy is due to the other induced even-frequency amplitudes. An A$_{\rm 2u}$ SC induces an even-orbital spin-singlet $s$-wave state, which gaps the spectrum. The A$_{\rm 1u}$ SC, on the other hand, induces an odd-orbital spin-triplet $s$-wave state, which does not gap the TI surface.\cite{Hao11} B$_{\rm 1u/2u}$ SCs induce even-orbital spin-singlet $d$-wave states, which gives a nodal quasiparticle spectrum. Finally, a spin-triplet E$_{\rm 2u}$ SC induces only odd-orbital spin-triplet $s$- and $d$-wave amplitudes, which do not influence the spectrum.
It is therefore crucial to know all induced even-frequency components for determining the energy spectrum of the superconducting TI surface state.
The LDOS of the odd-frequency amplitudes also have to be added to that of the conventional LDOS. However, the possible $\omega$-dependence of the odd-frequency even-orbital components only interferes with a fully gapped state, which is only found for the A$_{\rm 2u}$ SC where no odd-frequency intraorbital pairing exists. Odd-frequency odd-orbital pairing also never causes any sub-gap states.\cite{Black-Schaffer13new}

Since we use a full 3D model of the TI, and the results are valid even when the chemical potential is firmly situated within the bulk valence or conduction bands, Table I goes beyond surface proximity-induced superconductivity and displays, quite generally, the possible couplings between different pairing symmetries in TIs. Our results can thus also shed light on the recently discovered intrinsic superconducting state in Cu-doped Bi$_2$Se$_3$.\cite{Hor10} For example, Table I shows that the topological odd-orbital spin-triplet $s$-wave superconducting state proposed in Ref.~\onlinecite{Fu&Berg10} is present together with the A$_{\rm 1u}$ $p$-wave state. Other topological superconducting states discussed for Cu$_x$Bi$_2$Se$_3$\cite{Fu&Berg10, Hao11, Sasaki11} include odd-intraorbital pairing, which exists in conjunction with even-intraorbital $s$-wave pairing, and equal-spin $s$-wave pairing, which appears together with the $k_x+ik_y$ spin-triplet $p$-wave. The latter two topological states can thus be enhanced by proximity effect to a conventional $s$-wave SC or the  proposed $k_x+ik_y$ SC Sr$_2$RuO$_4$,\cite{Mackenzie03, Kallin12} respectively.

In summary, we have provided a full symmetry classification of the proximity-induced superconducting pairing amplitudes in a TI for spin-singlet $s$-wave, $d$-wave, and spin-triplet $p$-wave SCs. The Dirac surface state always gives rise to mixing between spin-singlet and spin-triplet states, as well as intraorbital odd-frequency pairing for spin-triplet SCs. We also find a complete interchangeability between odd-frequency and odd-orbital pairing because of the hybridized two-orbital nature of TIs. The different pairing amplitudes significantly modify the LDOS in the TI surface layer.

We are grateful to E. Abrahams for discussions and the Swedish and European research councils (VR, ERC) for funding.
Work at Los Alamos was supported by US DoE Basic Energy Sciences and in part by the Center for Integrated Nanotechnologies, operated by LANS, LLC, for the National Nuclear Security Administration of the U.S. Department of Energy under contract DE-AC52-06NA25396.


\end{document}